\documentclass[aps,prb,groupedaddress,showpacs,twocolumn]{revtex4}

\usepackage{graphicx}
\usepackage{amsmath}
\usepackage{bm}
\begin{document}
\renewcommand{\textfraction}{0.15}
\renewcommand{\topfraction}{0.85}

\title{Quasi-equilibrium optical nonlinearities in spin-polarized GaAs}
\author{Arjun Joshua}
\email{arjun@physics.iisc.ernet.in}
\author{V. Venkataraman}
\affiliation{Department of Physics, Indian Institute of Science\\
Bangalore, India 560012}


\begin{abstract}
Semiconductor Bloch equations, which microscopically describe the dynamics of a Coulomb interacting, spin-unpolarized electron-hole plasma, can be solved in two limits: the coherent and the quasi-equilibrium regime. These equations have been recently extended to include the spin degree of freedom, and used to explain spin dynamics in the coherent regime. In the quasi-equilibrium limit, one solves the Bethe-Salpeter equation in a two-band model to describe how optical absorption is affected by Coulomb interactions within a spin-unpolarized plasma of arbitrary density. In this work, we modified the solution of the Bethe-Salpeter equation to include spin-polarization and light holes in a three-band model, which allowed us to account for spin-polarized versions of many-body effects in absorption. The calculated absorption reproduced the spin-dependent, density-dependent and spectral trends observed in bulk GaAs at room temperature, in a recent pump-probe experiment with circularly polarized light. Hence our results may be useful in the microscopic modelling of density-dependent optical nonlinearities in spin-polarized semiconductors.
\end{abstract}

\maketitle

\section{Introduction}

As the nascent field of spintronics~\cite{sarmarmp} is merged with optoelectronics, it becomes increasingly important to understand the physics of the optical properties of spin-polarized semiconductors, that are nowadays studied by circularly polarized pump-probe experiments.~\cite{awschalombook} This degree of freedom of the spin, in the optical properties of photo-excited semiconductors, is mostly ignored in the literature. For spin-unpolarized photo-excited carriers in a pure semiconductor, it is known that the Coulomb attraction between the electron and hole created by a photon, causes excitonic resonances and enhances its absorption.~\cite{haugandkoch} 
As the background electron-hole plasma density is increased, the long-ranged Coulomb interaction between this photo-excited electron-hole pair and the plasma can no longer be neglected. Semiconductor Bloch equations (SBE) fully describe this interacting plasma and its time-dependent dynamics. SBE may be solved in two different time-scales: the coherent regime (where the induced dipole moment follows the optical field without de-phasing), or the quasi-equilibrium regime. In many applications, for example, semiconductor laser diodes, it may be assumed that the electrons and holes are thermalized within their respective bands.~\cite{haugpra} This quasi-equilibrium approximation simplifies the original coupled SBE. The resulting microscopic theory for the effect of a spin-unpolarized interacting plasma (arbitrary density) on absorption, agrees well with experiment.~\cite{lowenau} Here the linear optical susceptibility is got by solving the Bethe-Salpeter equation in the (quasistatic) screened ladder approximation.~\cite{lowenauprl,haugreview} In this formalism, the many-body effects of the plasma on absorption are the screening of the Coulomb enhancement, phase-space filling (PSF) by the carriers and bandgap renormalization (BGR). These effects can be viewed as density-dependent optical nonlinearities caused by the quasi-equilibrium plasma.~\cite{haugandkoch,haugpra} A two-band model is used for the conduction and valence bands by either ignoring the light hole ($lh$) band, or lumping it with the heavy hole ($hh$) band via an effective valence band density-of-states.~\cite{zimmermann}

Photo-excited carriers may be spin-polarized by the optical orientation technique.~\cite{oo} It is based on the selection rules for transitions induced by circularly polarized light, from both $hh$ and $lh$ bands into the conduction band. A three-band SBE to model intervalence band coherence of quantum wells, under circularly polarized photo-excitation was formulated,~\cite{binder} that included $hh$-$lh$ band coupling. For modelling optical response in spin-polarized bulk and quantum well semiconductors, a very general six-band SBE has been recently framed and applied (after neglecting some terms) to give a microscopic description of spin dynamics.~\cite{rossler,lechnerdp,lechnerbap} As was done for the (spin-unpolarized) semiconductor laser,~\cite{haugpra} it maybe desirable to solve such model spin-SBE in the quasi-equilibrium regime, for a microscopic understanding of a new type of spin optoelectronic device, the spin vertical cavity surface emitting laser (spin VCSEL).~\cite{rudolph} 

In the quasi-equilibrium regime, using the pump-probe technique Nemec et al.~\cite{nemec} recently studied absorption spectra in spin-polarized bulk GaAs at room temperature. They observed a spectral crossover in the difference in absorption between right ($\sigma^+$) and left ($\sigma^-$) circularly polarized light. This circular dichroism experienced by the probe, is due to the electronic spin-polarization excited by an earlier $\sigma^+$ pump pulse. For a microscopic description of this experiment, we present the spin-modified solution of the Bethe-Salpeter equation, extended to include the $lh$ band. Our approach is equivalent to solving the full spin-SBE in the quasi-equilibrium regime, but neglecting the terms corresponding to $hh$-$lh$ coupling, spin-splitting of the single particle states and electron-hole exchange interaction. The last two terms are important for spin relaxation processes, but not for optical transitions. $hh$-$lh$ coupling was neglected because it is important only for inter-valence band processes. Numerical simulations based on this framework, showed the spectral crossover and were in reasonable agreement with the experimental spin-dependent, density-dependent and spectral trends. The Bethe-Salpeter equation may therefore be useful in modelling spin-dependent many-body effects in semiconductors in the quasi-equilibrium regime.

\section{Method}\label{sec:method}
We first modify the solution of the Bethe-Salpeter equation, for spin-polarization ($\xi$) and inclusion of $lh$. Then the method used to compare with the time-dependent experimental data~\cite{nemec} is described.

\subsection{Spin-polarized optical susceptibility (including light holes)}

\begin{figure}[bhp]
\includegraphics[width=8cm, angle=0]{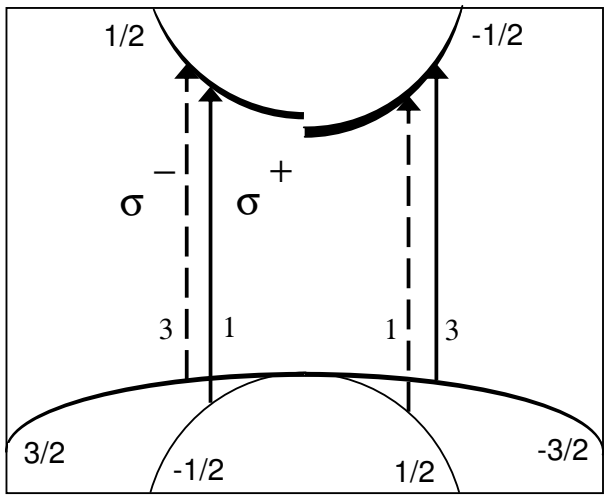}
\caption{\label{fig:selection}Selection rules in schematic band structure of bulk GaAs at the center of the Brillouin zone, with $n_\downarrow > n_\uparrow$. Incident light is assumed to be propagating in the $z$ direction. Bands are labelled with their $m_J$ indices, where $m_J$ is the component of total angular momentum $J$ along $z$ direction. Solid and dashed arrows, with the relative transition rates indicated at their base, correspond to $\sigma^+$ and $\sigma^-$ helicities of the probe. $\downarrow$ and $\uparrow$ electron bands are labelled with $m_J=-1/2$ and $1/2$ respectively. The spin-unpolarized valence bands are labelled with $m_J=\pm 3/2$ ($hh$) and $\pm 1/2$ ($lh$). PSF at a non-zero temperature is shown by the gradation in thickness of the bands (the small PSF of the $lh$ band has not been shown). The greater BGR of the $\downarrow$ band causes its bandedge to be lower compared to the $\uparrow$ band.}
\end{figure}

At $t=0$ both electron and hole spins are created by the right circularly polarized pump pulse in the experiment; however hole spins relax within $t\alt100$ fs leaving behind a spin polarization only from electrons. The selection rules favor creating spin-down ($\downarrow$) electrons from the $hh$ band three times as much as spin-up ($\uparrow$) electrons from the $lh$ band ($\downarrow$ and $\uparrow$ electrons have their spin opposite and along the propagation direction of the pump, respectively). Moreover the joint density-of-states effective mass for $hh$ transitions $m_{r,\,hh}$ is nearly twice $m_{r,\,lh}$ for $lh$ transitions. These two factors cause $n_\downarrow=6 n_\uparrow$, as pointed out in Ref.~\onlinecite{beck} ($n_\downarrow + n_\uparrow = n$, the plasma density). So, at $t=0$ the spin polarization $\xi_{\text{max}}=5/7$, where $\xi$ is defined as
\[  \xi=\frac{n_\downarrow - n_\uparrow}{n_\downarrow + n_\uparrow} . \]
Later the same selection rules involving the $hh$ and $lh$ bands, determine the absorption of the right ($\sigma^+$) or left ($\sigma^-$) circularly polarized probe (Fig.~\ref{fig:selection}). Therefore, here the absorption depends not just on the total density of electrons and holes (as described in Refs.~\onlinecite{haugandkoch,haugpra,lowenau,lowenauprl,haugreview,zimmermann} for $\xi=0$) but individually on $n_\downarrow$, $n_\uparrow$, $n_{hh}$ and $n_{lh}$.

PSF reduces absorption since lesser number of states are made available for optical transitions. BGR increases absorption since the transitions take place at larger wavevector, as the bandgap is narrowed. PSF and BGR, which are density-dependent, are different for transitions into the $\downarrow$ and $\uparrow$ electron bands (Fig.~\ref{fig:selection}). As will be seen later, these transitions also have different Coulomb enhancement because the enhancement itself depends on PSF and BGR. All this affects the susceptibility $\chi^{cv}(\omega)$ pertaining to optical transitions from the valence ($v=hh,\,lh$) to the conduction ($c=\downarrow,\,\uparrow$) bands. The susceptibility $\chi^{\pm}(\omega)$ for $\sigma^{\pm}$ can be written as a sum of two such transitions (Fig.~\ref{fig:selection}),
\begin{subequations}\label{eq:selection}
\begin{eqnarray}
\chi^+(\omega)&=& \chi^{\downarrow,\,hh}(\omega) + \chi^{\uparrow,\,lh}(\omega) \\
\chi^-(\omega)&=& \chi^{\uparrow,\,hh}(\omega) + \chi^{\downarrow,\,lh}(\omega).
\end{eqnarray}
\end{subequations}
$\chi^+$ is sensitive to the $\downarrow$ band, whereas $\chi^-$ is weighted towards the $\uparrow$ band, because the transition from the $hh$ band is favored over that from the $lh$ band. Therefore, in general a circular dichroism results from a non-zero $\xi$. The complex optical dielectric function $\epsilon^{\pm}(\omega)$ is got from $\chi^{\pm}(\omega)$ using
\begin{equation}
\epsilon^{\pm}(\omega)=\epsilon_{\infty} + 4\pi\chi^{\pm}(\omega), \label{eq:dielectric}
\end{equation}
from which the absorption $\alpha^{\pm}(\omega)$ and circular dichroism $\Delta\alpha(\omega)=\alpha^+(\omega) - \alpha^-(\omega)$ were obtained. Also, since linearly polarized light can be written in terms of $\sigma^+$ and $\sigma^-$, we obtain
\begin{equation}
\chi^0=\frac{\chi^+ + \chi^-}{2},
\end{equation}
where $\chi^0$ is the susceptibility of linearly polarized light. It may be verified that if $\xi=0$, $\chi^0=\chi^+=\chi^-$.

The optical susceptibility $\chi^{cv}(\omega)$ of a particular transition was related to the microscopic susceptibility $\chi^{cv}(k,\omega)$ by
\begin{equation}
\chi^{cv}(\omega) = \frac{1}{L^3} \sum_{\bm{k}} d^{cv}(k) \chi^{cv}(k,\omega). \label{eq:chiomega}
\end{equation}
The sum over wavevector does \textit{not} include the spin degeneracy. The interband dipole matrix element for circularly polarized light, $d^{cv}(k)$, reflected the selection rules,
\begin{subequations}
\begin{eqnarray}
d^{cv}(k) &=& \frac{e\hbar P_0}{m_0 E_g\left( 1+\frac{\hbar^2 k^2}{2 m_{r,v} E_g} \right)}\;\;\;(v=hh) \label{eq:dcv}   \\
 &=& \sqrt{\frac{1}{3}}\,\frac{e\hbar P_0}{m_0 E_g\left( 1+\frac{\hbar^2 k^2}{2 m_{r,v} E_g} \right)}\;\;\;(v=lh),
\end{eqnarray}
\end{subequations}
where $m_0$ is the free electron mass, $m_{r,\,v}=(1/m_e+1/m_v)^{-1}$ and $E_g$ is the bandgap. The momentum matrix element $P_0$, in an eight-band model~\cite{supriyodatta} is given by
\begin{equation}
P_0^2 \simeq \frac{m_0 E_g}{2} \left( \frac{m_0}{m_e}-1 \right) \frac{3E_g+3\Delta_{so}}{3E_g+2\Delta_{so}},
\end{equation}
where $\Delta_{so}$ is the spin-orbit splitting of the valence band at the center of the Brillouin zone.

\subsubsection{Interacting optical spectrum}
We repeat the steps for getting $\chi^{cv}(\omega)$ outlined in Ref.~\onlinecite{ell}, but with modifications for $lh$ and spin. $\chi^{cv}(k,\omega)$, which causes the dichroism, is got by numerically solving the effective Bethe-Salpeter equation describing the repetitive electron-hole scattering (ladder approximation):
\begin{widetext}
\begin{equation}
\left[ \hbar\omega-E_g-\Delta E_g^{cv}+i\,\gamma^{cv}(k,\omega) \right] \chi^{cv}(k,\omega) = -(1-f_c(k)-f_v(k)) \left[ d^{cv}(k) + \sum_{\bm{k'}}V_s(|\bm{k}-\bm{k'}|) \chi^{cv}(k',\omega) \right].\label{eq:bse}
\end{equation}
\end{widetext}
The screened Coulomb potential $V_s(q)$ derived in the random phase approximation, and simplified in the quasistatic single plasmon-pole approximation is given by
\begin{eqnarray}
V_s(q) &=& \frac{1}{L^3} \frac{4\pi e^2}{\epsilon_0 q^2} \left[1 - \frac{1}{1+\frac{q^2}{\kappa^2} + \left(\frac{\nu_q}{\omega_{pl}} \right)^2 } \right], \label{eq:potential} \\
\omega_{pl}^2 &=& \frac{4\pi e^2}{\epsilon_0} \sum_j \frac{n_j}{m_j},\;\;\;(j=\downarrow,\,\uparrow,\,hh,\,lh) \\
\kappa^2 &=& \frac{4\pi e^2}{\epsilon_0} \sum_j \frac{\partial n_j}{\partial \mu_j},   \label{eq:wavenumber}
\end{eqnarray}
where $\mu_j$ is the chemical potential, $\omega_{pl}$ and $\kappa$ are the 3$d$ plasma frequency and wave number respectively and $\nu_q^2$ simulates the electron-pair continuum. Note that $\omega_{pl}$ and $\kappa$ have been made $\xi$-dependent. In Eq.~(\ref{eq:bse}), the factor $(1-f_c(k)-f_v(k))$ causes PSF. The Fermi functions $f_j(k)$ that described the distribution of the electrons and holes were
\begin{equation}
f_j(k)=\frac{1}{\exp[\beta(E_j(k)-\mu_j)] + 1}
\end{equation}
where $E_j(k)=\hbar^2 k^2/(2 m_j)$ and $\beta=1/(k_B T)$. We ignored the small spin-dependence of the electron mass~\cite{zhang} (i.e. $m_\downarrow=m_\uparrow=m_e$). The BGR, $\Delta E_g^{cv}$, for an optical transition between the valence band and the conduction band is
\begin{equation}
\Delta E_g^{cv}(k)=e_c(k)+e_v(k).
\end{equation}
The self-energy $e_j(k)$ of the $j^{\text{th}}$ quasiparticle, in the quasistatic approximation is
\begin{equation}
e_j(k) \simeq - \sum_{\bm{k'}}V_s(|\bm{k}-\bm{k'}|) f_j(k') + \frac{1}{2} [V_s(r=0)-V(r=0)],
\end{equation}
where the first term is the `screened-exchange' and the second term is the `Coulomb-hole'. The Coulomb-hole term is the same for each of the $j$ quasiparticles. In Eq.~(\ref{eq:potential}) $\nu_q^2$, which simulates the electron-pair continuum, as well as the temperature-dependent damping $\gamma^{cv}(k,\omega)$ and bandgap $E_g$ in Eq.~(\ref{eq:bse}) were got from Refs.~\onlinecite{ell} and~\onlinecite{lowenau}. The chemical potentials $\mu_j$, assuming that the electrons and holes were in quasi-equilibrium, was got using a form of the Aguilera-Navarro approximation.~\cite{ell} The $hh$ and $lh$ have the same chemical potential because they are in equilibrium with each other.

Rearranging Eq.~(\ref{eq:bse}) we got
\begin{equation}
\chi^{cv}(k,\omega)=\chi_0^{cv}(k,\omega)\left[ 1 + \frac{1}{d^{cv}(k)}\sum_{\bm{k'}}V_s(|\bm{k}-\bm{k'}|) \chi^{cv}(k',\omega) \right],\label{eq:chibse}
\end{equation}
where
\begin{equation}
\chi_0^{cv}(k,\omega)=- \frac{d^{cv}(k) (1-f_c(k)-f_v(k))}{\hbar\omega - E_g-\Delta E_g^{cv} +i\,\gamma^{cv}(k,\omega)}. \label{eq:chi0}
\end{equation}
To get the correct crossover between gain and absorption with $\gamma^{cv}(k,\omega)$, $\chi_0^{cv}(k,\omega)$ was described by a spectral representation.~\cite{ell} By defining
\begin{equation}
\chi^{cv}(k,\omega)=\Gamma^{cv}(k,\omega) \chi_0^{cv}(k,\omega), \label{eq:gamma}
\end{equation}
and substituting it into Eq.~(\ref{eq:chibse}), we obtained for the vertex function $\Gamma^{cv}(k,\omega)$,
\begin{equation}
\Gamma^{cv}(k,\omega)=1 + \frac{1}{d^{cv}(k)}\sum_{\bm{k'}}\bar{V}_s(k,k') \chi_0^{cv}(k',\omega) \Gamma^{cv}(k',\omega).\label{eq:vertex}
\end{equation}
$V_s(|\bm{k}-\bm{k'}|)$ has been replaced in Eq.~(\ref{eq:vertex}) by its angle-averaged value $\bar{V}_s(k,k')$, because we assumed that only $s$-wave scattering contributed to the optical transitions. We used a matrix of approximately $200\times 200$ Gauss-Legendre quadrature points to represent the vertex integral equation~(\ref{eq:vertex}). The diagonal singularity of the matrix was regularized by the compensation technique,~\cite{haugreview} before it was inverted to give the solution for $\Gamma^{cv}(k,\omega)$. The optical susceptibility $\chi^{cv}(\omega)$ of a particular transition was obtained from
\begin{equation}
\chi^{cv}(\omega) = \frac{1}{L^3} \sum_{\bm{k}} d^{cv}(k) \Gamma^{cv}(k,\omega) \chi_0^{cv}(k,\omega), \label{eq:chiomega2}
\end{equation}
where we substituted Eq.~(\ref{eq:gamma}) in Eq.~(\ref{eq:chiomega}).

\subsubsection{Noninteracting optical spectrum}
If $\Gamma^{cv}(k,\omega)$ is neglected in Eq.~(\ref{eq:chiomega2}) we obtain the `noninteracting' susceptibility, which differs from the truly noninteracting susceptibility due to the BGR term present in $\chi_0^{cv}(k,\omega)$ [c.f. Eq.~(\ref{eq:chi0})]. $\Gamma^{cv}(k,\omega)$, which depends on PSF and BGR [c.f. Eq.~(\ref{eq:vertex}) and~(\ref{eq:chi0})], causes the excitonic resonances and the Coulomb enhancement. This is because $\Gamma^{cv}(k,\omega)$ expresses the influence of multiple electron-hole scattering (resulting from their attractive interaction), on the susceptibility.

It is difficult to separate PSF and BGR in the interacting $\Delta\alpha(\omega)$ due to the presence of the $\Gamma^{cv}(k,\omega)$ term. Therefore we used the noninteracting $\chi^{cv}(\omega)$ to study how competition between PSF and BGR influences $\Delta\alpha(\omega)$.

In our calculated spectra, we took the spectral representation for only the imaginary part of $\chi_0^{cv}(k,\omega)$.~\cite{ell} Also, the $k$-dependent BGR was taken as a rigid shift at $k_F=\left( 3 \pi^2 n \right)^{1/3}$. These simplifications are not expected to significantly affect the results. The material parameters used were: exciton Rydberg $E_0=4.2\,\text{meV}$, exciton Bohr radius $a_0=125$ \AA, electron mass $m_e=0.0665\,m_0$, $hh$ mass $m_{hh}=0.457\,m_0$, $lh$ mass $m_{lh}=0.08\,m_0$, $\Delta_{so}=0.341\,\text{eV}$ $\epsilon_0=13.71$, $\epsilon_{\infty}=10.9$ and $T=295\,\text{K}$.

\subsection{Comparison with experiment}
The probe spectral width was accounted for by adding its `half-width at half-maximum' value of 15 meV to $\gamma^{cv}(k,\omega)$ in the calculation. The experimental data is mostly in terms of the normalized differential transmittance $D$
\begin{equation}
D=\frac{\left( \frac{\Delta T}{T} \right)^+ - \left( \frac{\Delta T}{T} \right)^- }{\left( \frac{\Delta T}{T} \right)^+ + \left( \frac{\Delta T}{T} \right)^- }.
\end{equation}
Here $(\Delta T/T)^{\pm}=(T^{\pm}-T)/T$, where $T^{\pm}$ is the transmission of probe $\sigma^{\pm}$ after the sample is pumped with $\sigma^+$. $T$ is the unpumped transmission through the bulk sample of thickness $w\approx 1\mu\text{m}$. If the change from the unpumped absorption $\Delta\alpha^{\pm}$ is such that $\Delta\alpha^{\pm} \ll 1/w \approx 10^4 \text{cm}^{-1}$, we can write $D$ as
\begin{eqnarray}
D &\simeq& \frac{\Delta\alpha^+ - \Delta\alpha^-}{\Delta\alpha^+ + \Delta\alpha^-} \label{eq:D} \\
  &=& \frac{\alpha^+ - \alpha^-}{\alpha^+ + \alpha^- - 2\alpha_0} = \frac{\Delta\alpha}{\alpha^+ + \alpha^- - 2\alpha_0}.
\end{eqnarray}
The unpumped absorption $\alpha_0(\omega)$ could be calculated from $\chi^0(\omega)$ with $\xi=0$ and the background doping density of the sample,~\cite{nemec} $n_0=10^{15}\,\text{cm}^{-3}$. The sign of $D$ is opposite to that of $\Delta\alpha$ because screening by the pumped carriers usually reduces the absorption i.e. $\alpha^+ + \alpha^- - 2\alpha_0$ is a negative quantity. Furthermore, $D$ is not affected by an overall density-independent scaling factor $C(\omega)$ that could multiply the calculated absorption, since $D$ is a normalized quantity. Usually $C(\omega)$ is needed to match the calculated absorption to the experimental value of absorption in a `pure', unexcited sample.~\cite{zimmermann}

The  $\xi$-dependent calculations were compared with the time-dependent experimental data by using the following equation describing spin relaxation,
\begin{equation}
\xi(t)=\xi_{\text{max}}\,e^{-2t/\tau_s},
\end{equation}
where $\xi_{\text{max}}=5/7$ and the spin relaxation time~\cite{nemec} $\tau_s = 130\,\text{ps}$. We assumed a temporally constant plasma density $n$ since $t\ll \tau_r\approx 1\,\text{ns}$, where $\tau_r$ is the carrier recombination time.

\section{Results and Discussion}
\begin{figure}[bhp]
\includegraphics[width=8.6cm, angle=0]{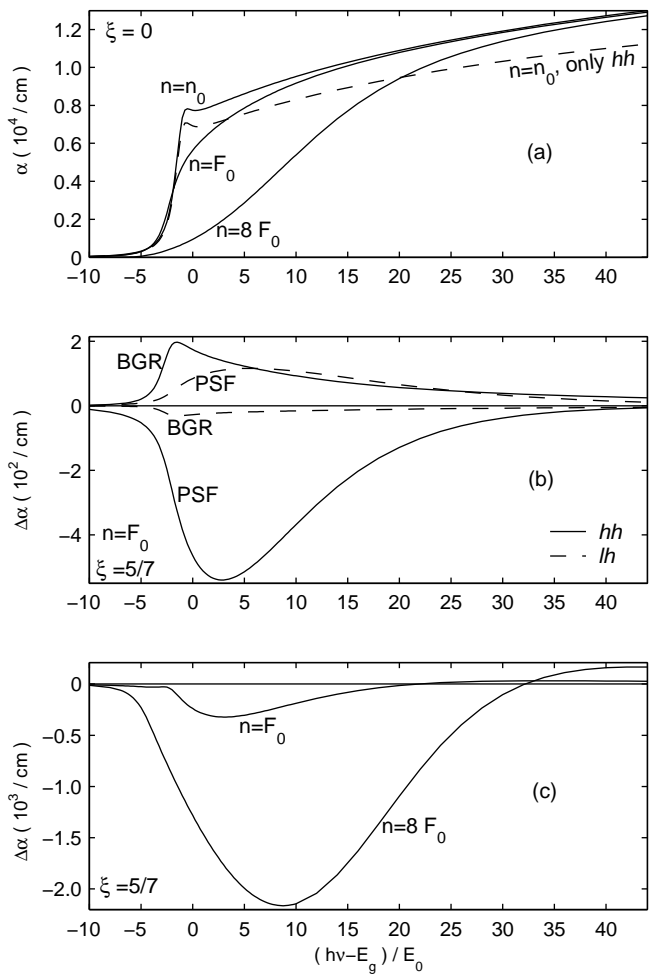}
\caption{\label{fig:intro}(a) Spin-unpolarized interacting $\alpha$, with background $n_0=10^{15}$ cm$^{-3}$ and photo-excited plasma densities $F_0=1.3\times 10^{17}$ cm$^{-3}$ and $8 F_0$. The dashed curve differs from the solid curve with $n=n_0$, by the neglect of $lh$. (b) Noninteracting $\Delta\alpha$ decomposed into $hh$ and $lh$ transitions, each of which has contributions from PSF and BGR. Within a transition, PSF and BGR oppose each other. Between transitions, PSF and BGR from $hh$ oppose those from $lh$. (c) Density-dependence of total noninteracting $\Delta\alpha$. The curve for $n=F_0$ can be got by summing together the curves in (b).}
\end{figure}

We show that our calculation captures all the trends in the experimental data, in probed energy $h\nu$, spin polarization $\xi$ and pumped density $n$. To explain how PSF and BGR cause a ($\xi$-independent) crossover in $\Delta\alpha$, we first discuss results for noninteracting $\Delta\alpha$. After showing the effect of the Coulomb interaction on $\Delta\alpha$, we finally compare the results for interacting $\Delta\alpha$ with the data. The agreement with the data comes directly from our calculation, without requiring us to adjust $\xi$, $n$ or sample-dependent broadening.

We did two checks on the spin-polarized calculation. Putting $\xi=0$ gave back the unpolarized absorption (Fig.~\ref{fig:intro}(a)) as calculated by Ref.~\onlinecite{ell}, for different plasma densities. At lower densities, with $\xi=0$, we also got the broadened Elliott's formula~\cite{goni} by removing the $k$-dependence of $\gamma^{cv}(k,\omega)$ and $d^{cv}(k)$. The discrepancy with Elliott's formula was $\approx 10^{-3}$ for photon energies that exceeded the bandgap.

Inclusion of $lh$ serves only to enhance the unpolarized absorption by 1/6, 1/3 from the matrix element and 1/2 from the density-of-states (compare dashed and solid curves for $n=n_0$ in Fig.~\ref{fig:intro}(a)). Treating $lh$ independently (without $hh$-$lh$ coupling) did not lead to noticeable artifacts in the calculated absorption.

\subsection{Noninteracting optical spectra}
The microscopic noninteracting calculation agrees with the earlier simplified explanation~\cite{nemec} that the $\Delta\alpha$ crossover is caused by competition between PSF and BGR. Within a transition (either $hh$ or $lh$), PSF dominated at lower energies whereas BGR (of the opposite sign) dominated at higher energies. Moreover, PSF and BGR from $lh$ transitions were opposite in sign (due to the selection rules in Eq.~(\ref{eq:selection})) compared to those from $hh$ transitions (Fig.~\ref{fig:intro}(b)). The $\Delta\alpha$ crossover shifted to higher energies as $n$ was increased, because PSF became more important~\cite{nemec} (Fig.~\ref{fig:intro}(c)).

\begin{figure}[bhp]
\includegraphics[width=8.6cm, angle=0]{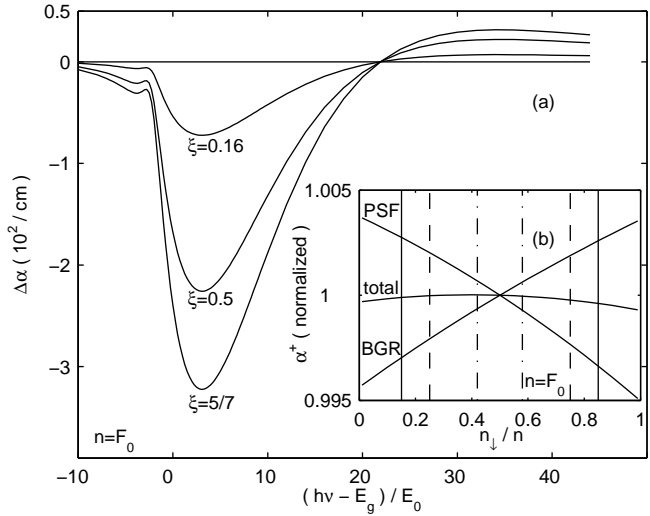}
\caption{\label{fig:spindep}(a) $\xi$-dependence of noninteracting $\Delta\alpha$. (b) PSF and BGR components of noninteracting $\alpha^+$ at the crossover energy vs. $n_\downarrow$ (neglecting $lh$ band). The ordinate is normalized to the value of $\alpha^+$ at $n_\downarrow=n/2$, whereas the abscissa is normalized to $n$. The opposite $n_\downarrow$-dependence of PSF and BGR add up to give total $\alpha^+$ (scaled by 0.5) that is almost $n_\downarrow$-independent. The $\xi$-independence of $\Delta\alpha$ at the crossover energy in (a) may be got from the difference of the value of total $\alpha^+$ at the $n_\downarrow$ values shown by vertical solid lines ($\xi=5/7$), dashed lines ($\xi=0.5$) and dot-dashed lines ($\xi=0.16$).}
\end{figure}

In contrast, the $\Delta\alpha$ crossover energy was independent of $\xi$, at a fixed $n$ (Fig.~\ref{fig:spindep}(a)), due to the PSF and BGR interplay. A similar $\xi$-independent crossover occurred at higher energies, if $hh$ transitions alone were considered. The weaker $lh$ transitions did not result in a crossover in $\Delta\alpha$ in the energy interval shown, as can be verified by summing the dashed curves in Fig.~\ref{fig:intro}(b). Since $\xi$ determines $n_\downarrow$ relative to $n_\uparrow$, we study the cause of the $\xi$-independent crossover by plotting $\alpha^+$ (inducing transitions only between the $hh$ and $\downarrow$ bands after neglecting the $lh$ transitions) vs. $n_\downarrow$ (Fig.~\ref{fig:spindep}(b)). Analogous results were got by plotting $\alpha^-$ vs. $n_\uparrow$. The $n_\downarrow$-independence of total $\alpha^+$ at the crossover energy, indicates that opposing trends between PSF and BGR for each helicity of light, caused the $\Delta\alpha$ crossover to be independent of $\xi$. Only the electronic part of the $lh$ transitions contributed to $\Delta\alpha$ (holes are unpolarized), hence the $\xi$-independent nature of the crossover was unaffected in the full $\Delta\alpha$ which included $lh$ transitions (Fig.~\ref{fig:spindep}(a)).

\subsection{Interacting optical spectra}

\begin{figure}[bhp]
\includegraphics[width=8.6cm, angle=0]{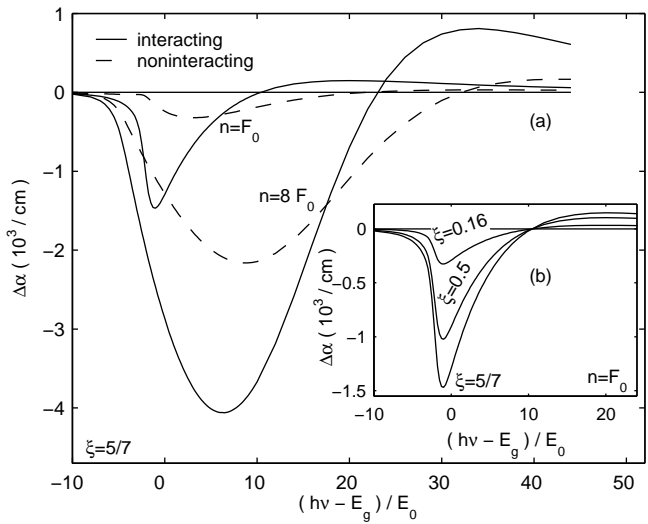}
\caption{\label{fig:interact} (a) Density-dependence of interacting $\Delta\alpha$. The corresponding noninteracting $\Delta\alpha$ shown, is identical to Fig.~\ref{fig:intro}(c). (b) $\xi$-dependence of interacting $\Delta\alpha$.}
\end{figure}

The Coulomb interaction increased the magnitude of $\Delta\alpha$ and shifted its crossover energy, but did not affect the trends in $h\nu$, $n$ and $\xi$. The interaction caused a sharp peak near the bandgap (due to excitonic effects) in the interacting $\Delta\alpha$, for $n=F_0$, and also enhanced it at higher energies compared to the noninteracting $\Delta\alpha$ (compare solid and dashed curves for $n=F_0$ in Fig.~\ref{fig:interact}(a)). At $n=8 F_0$, the peak near the bandgap in the interacting $\Delta\alpha$ was not as pronounced (due to the almost complete excitonic ionization) compared to the noninteracting case (solid and dashed curves for $n=8F_0$ in Fig.~\ref{fig:interact}(a)). But at higher energies, they show a persisting Coulomb enhancement even at $n=8F_0$. This is because the screening by the plasma is not very effective at high energies~\cite{zimmermann} (solid curves for $n=n_0$ and $8 F_0$ in Fig.~\ref{fig:intro}(a) approach each other at high energies). However the $\xi$-independence of the $\Delta\alpha$ crossover energy was preserved, despite Coulomb interactions (Fig.~\ref{fig:interact}(b)). The cause of the shift of the crossover to lower energies when we compare the interacting with the noninteracting curves is unclear.

\begin{figure}[bhp]
\includegraphics[width=8.6cm, angle=0]{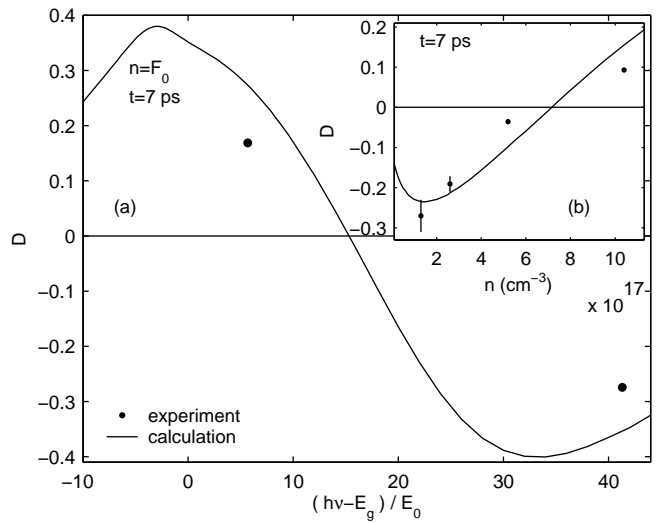}
\caption{\label{fig:exp_spectrum} Comparison with experiment. (a) Spectral $D$. The data points are got from Figs. 4(b) and 6(b) of Ref.~\onlinecite{nemec}, at $t=7$ ps. (b) Density-dependence of $D$ at a fixed probe energy. The experimental points are for $n=F_0$, $2 F_0$, $4 F_0$ and $8F_0$, again at $t=7$ ps (from Fig 7 of Ref.~\onlinecite{nemec}). The calculation used $\lambda=816$ nm whereas experimentally $\lambda=775$ nm. Vertical bars indicate the experimental noise.}
\end{figure}

\subsection{Comparison with experiment}
We compare our calculation with the experiment by accounting for probe width in the broadening (Fig.~\ref{fig:exp_spectrum}(a)). We think that spectral averaging due to this additional broadening causes the crossover to shift from $(h\nu-E_g)/E_0=10.5$ (solid curve for $n=F_0$ in Fig.~\ref{fig:interact}(a)) to $(h\nu-E_g)/E_0=15.3$. Because $D$ is plotted instead of $\Delta\alpha$ in Fig.~\ref{fig:exp_spectrum}(a), there is an overall flip of sign compared to Fig.~\ref{fig:interact}. 

\begin{figure}[bhp]
\includegraphics[width=8.6cm, angle=0]{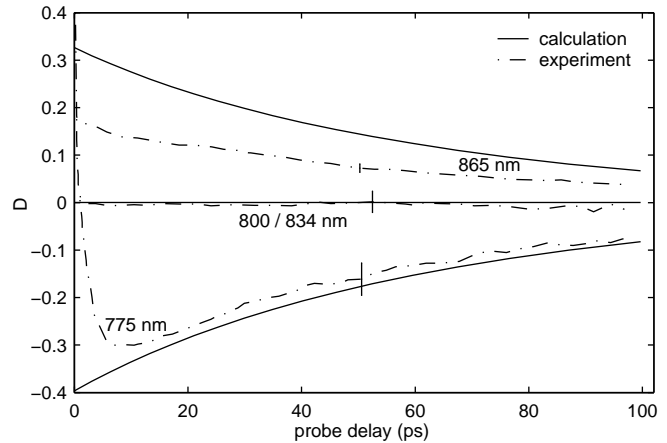}
\caption{\label{fig:exp_wavedep} Comparison with experiment (dot-dashed lines) regarding evolution of $D$ at the probe $\lambda=775$ nm, 800 nm (834 nm in the calculation) and 865 nm. The experimental data are got from Fig. 8 of Ref.~\onlinecite{nemec}. Pumped densities are $1.3\times10^{17}$ cm$^{-3}$, $6\times10^{16}$ cm$^{-3}$ and $2\times10^{16}$ cm$^{-3}$ respectively.~\cite{nemec} Vertical bars indicate the experimental noise.}
\end{figure}

Our model reproduces the experimental observations that $D$ changes sign either as $\lambda$ is varied at a fixed $n$ (Fig.~\ref{fig:exp_spectrum}(a)) or $n$ is varied at a fixed $\lambda$ (Fig.~\ref{fig:exp_spectrum}(b)). Furthermore, the expected $\xi$-independence of the crossover energy (Fig.~\ref{fig:interact}(b)) is indeed shown by the data (flat dashed curve in Fig.~\ref{fig:exp_wavedep}). In Figs.~\ref{fig:exp_spectrum}(b) and~\ref{fig:exp_wavedep} the calculation used $\lambda=816$ nm and $\lambda=834$ nm instead of the actual probe $\lambda=775$ nm and $\lambda=800$ nm respectively. The tendency of $D$ to change sign at a fixed probe wavelength or energy (Fig.~\ref{fig:exp_spectrum}(b)) as $n$ is increased, occurred over a certain range of energies. This tendency is also shown by $\Delta\alpha$ over $10.5<(h\nu-E_g)/E_0<23.1$ (solid curves for $n=F_0$ and $8F_0$ in Fig.~\ref{fig:interact}(a)). Outside this range $|D|$ increased, without flipping its sign, with $n$. We cannot obtain the carrier thermalization part of the experimental spectra ($t<1$ps) in Fig.~\ref{fig:exp_wavedep} because it has not been accounted for in the calculation.

The maximum value of $D$ due to PSF is 0.35 (instead of 0.25 as stated in Ref.~\onlinecite{nemec}). This is because, by assuming $\Delta\alpha \sim -n$ in Eq.~(\ref{eq:D}), we obtain $D=\xi/2$ and as noted in Section~\ref{sec:method}, $\xi_{\text{max}}=5/7$. The numerically calculated $D$ supported the above reasoning: in Figs.~\ref{fig:exp_spectrum} and \ref{fig:exp_wavedep}, $D\alt 0.35$ at lower energies. However near the bandgap, perhaps the excitonic nonlinearity slightly increased the value of $D$ (Fig.~\ref{fig:exp_spectrum}(a)). $|D|$ significantly exceeds 0.35 only at higher energies where PSF is expected to be less important (since $D$ is negative). The quantitative mismatch with experiment, at higher energies may be due to the neglect of dynamic screening and band non-parabolicity. We adjusted within a factor of 2 the values of $\nu_q^2$ and the numerical constant $\alpha$ (describing $\gamma(k,\omega)$),~\cite{ell} but found that the crossover energy changed by $\alt 1\%$. Experimental uncertainties in the photo-excited carrier density $n$ and the induced spin-polarization $\xi$ as well as sample-dependent broadening, none of which were adjusted for, are also expected to affect the mismatch.

\section{Conclusion}
We have provided a method to calculate spin-polarized many-body effects in the room temperature absorption spectra of bulk GaAs. This was done by modifying the existing microscopic theory for absorption, that is known to match experiment over a wide range of plasma densities. Light hole contributions were also included. The agreement with results of a recent circularly polarized pump-probe experiment came directly from our calculation, without requiring us to adjust spin-polarization, plasma density or sample-dependent broadening. We find that the $\Delta\alpha$ crossover and the experimental spin-dependent, density-dependent and spectral features are reproduced, thus validating the use of our model to understand circularly polarized pump-probe experiments in III-V semiconductors, in the quasi-equilibrium regime. This also opens up the possibility that the Bethe-Salpeter equation may be used to theoretically describe spin-dependent many-body nonlinearities in the operation of the spin VCSEL and perhaps make predictions about its performance.

\begin{acknowledgments}
We thank the Department of Science and Technology, Government of India, for partial financial support.
\end{acknowledgments}

\bibliography{arjun}

\end{document}